# International Co-authorship Relations in the *Social Science Citation Index*: Is Internationalization Leading the Network?

*Journal of the American Society for Information Science and Technology* (in press)

Loet Leydesdorff,[a]* Han Woo Park,[b] & Caroline Wagner[c]

**Abstract**

International co-authorship relations have increasingly shaped another dynamic in the natural and life sciences during recent decades. However, much less is known about such internationalization in the social sciences. In this study, we analyze international and domestic co-authorship relations of all citable items in the DVD-version of the Social Science Citation Index 2011 (SSCI). Network statistics indicate four groups of nations: (*i*) an Asian-Pacific one to which all Anglo-Saxon nations (including the UK and Ireland) are attributed; (*ii*) a continental European one including also the Latin-American countries; (*iii*) the Scandinavian nations; and (*iv*) a community of African nations. Within the EU-28, eleven of the EU-15 states have dominant positions. In many respects, the network parameters are not so different from the Science Citation Index. In addition to these descriptive statistics, we address the question of the relative weights of the international versus domestic networks. An information-theoretical test is proposed at the level of organizational addresses within each nation; the results are mixed, but the international dimension is more important than the national one in the aggregated sets (as in the Science Citation Index). In some countries (e.g., France), however, the national distribution is leading more than the international one. Decomposition of the USA in terms of states shows a similarly mixed result; more US states are domestically oriented in SSCI, whereas more internationally in SCI. The international networks have grown during the last decades in addition to the national ones, but not by replacing them.

**Keywords:** co-authorship, international, domestic, map, European Union, network

[a] Amsterdam School of Communication Research (ASCoR), University of Amsterdam, Kloveniersburgwal 48, 1012 CX Amsterdam, The Netherlands; loet@leydesdorff.net ; http://www.leydesdorff.net; * corresponding author
[b] Department of Media & Communication, YeungNam University, 214-1, Dae-dong, Gyeongsan-si, Gyeongsangbuk-do, South Korea, 712-749; hanpark@ynu.ac.kr.
[c] John Glenn School of Public Affairs, The Ohio State University, Columbus, OH 43210, USA; cswagner@mac.com .

# 1. Introduction

The network of international co-authorship relations has been intensively researched by scientometricians because of its spectacular increase in recent decades (e.g., Glänzel, 2001; Luukkonen *et al.*, 1993; Melin & Persson, 1996; Okubo *et al.*, 1992; Persson *et al.*, 2004; Wagner & Leydesdorff, 2005). The *Science and Engineering Indicators* 2012 of the National Science Board of the USA (NSB, 2012), for example, devotes an entire section (at pp. 5-35 ff.) to "Co-authorship and Collaboration." Figure 5-25 in this report shows the long-term trends in co-authorship relations, both internationally and for US papers.

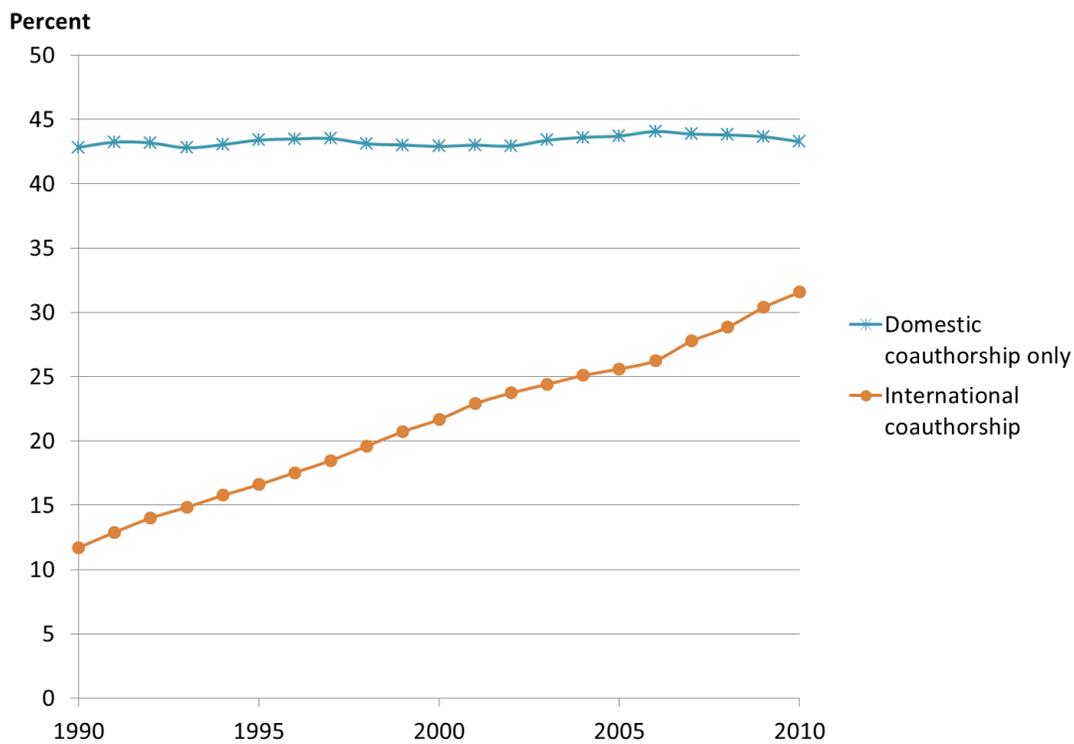

**Figure 1**: National and international coauthorships of US-based articles; counts from sets of journals covered by the Science Citation Index (SCI) and Social Science Citation Index (SSCI); whole-number counted. Source: NSB, 2012, at p. 5-37.

Figure 1 shows an adaptation of the part of this figure (Fig. 5-25) that focuses on co-authored articles with at least one address in the USA. For the rest of the world, the trends cannot be so neatly separated because domestic co-authorship has also increased, but the trends in increases in the development of international co-authorship relations are worldwide and show very similar patterns by region and by field of science (e.g., Adams *et al.*, in press; Leydesdorff *et al.*, 2013; Wagner, 2005).

The rapid and sustained rise in international collaboration evidenced by co-authorships suggests that the underlying social organization involved in creating scientific knowledge has been shifting. This shift has important implications for management and policy, since most public funds are invested with the goal of building national capacities. But, scientific activities are self-organizing in response to the frontiers of knowledge, and scientists seek collaborators who can help advance their research—whether they are co-nationals or not. Meantime, the institutions that coordinate and fund these activities are not so easily adaptable: institutions governing science can be expected to change at a slower rate than the activities that they manage. They are challenged by the rapid rate of change within the science system towards collaboration, internationalization, and interdisciplinary linkages. These changes raise significant questions for the social and political supporting institutions. A different dynamics in the network of international co-authorship relations may lead to spill-overs of national (or regional) investments (Wagner, 2008).

In previous studies, we have asked these questions with reference to the Science Citation Index (e.g., Leydesdorff & Wagner, 2008; Leydesdorff, Wagner, Park, and Adams, 2013) or at the level of specific disciplines (such as Library and Information Science; Leydesdorff & Persson, 2010). Leydesdorff, Hammarfelt, and Salah (2010) explored the disciplinary composition of the Arts & Humanities Citation Index based on a download of the 2008-volume. Social sciences can be expected to operate differently at the level of social organization for several reasons.

Social sciences have been shown to have distinct citing patterns that differ from the natural and engineering sciences (e.g., Nederhof, 2006). Social scientists are more likely to cite deep into history while the natural and engineering sciences are more likely to have citations at the research front (Price, 1970). Social sciences are generally not as well funded by the national-level research organizations that fund science: they claim a much smaller share of total funds. As a result, social sciences are more likely to be funded locally, and therefore the work may very well focus on more local or even idiosyncratic problems. Perhaps most importantly, natural and engineering sciences (while they vary among themselves) seek to create knowledge that is verifiable independent of cultural influences, where social sciences may be heavily invested in cultural variability.

This study was occasioned by the opportunity to study the complete Social Science Citation Index 2011 (SSCI) as distinct from the Science Citation Index 2011 (SCI) using the DVD-versions of the two databases. The two databases are available as separate DVDs, but the focus of most studies is on the SCI or—as in the case of the *Science and Engineering Indicators*—on the two databases combined.[1] We have kept the two datasets apart and downloaded both sets at the document level. The separate versions allow us to draw networks from SSCI and compare them to SCI networks[2] to examine the extent of international collaboration and the influence of that collaboration on geographically proximate research. In the descriptive part, we make comparisons with equivalent statistics for SCI.

---

[1] This data is compiled by PatentBoard™ under a contract of the National Science Foundation of the USA.

We expected to find important differences between networks from SCI and SSCI based on different underlying social dynamics (e.g., Larivière *et al.*, 2006; Ossenblok *et al.*, 2012; Ossenblok *et al.*, in press). We expected social sciences to be more locally focused and therefore to be less internationalized than SCI. We also expected to find that national-level influences will be more important in SSCI than in SCI. The social sciences do not have the equivalent of the "big sciences" such as physics which rely heavily on equipment, so we expected the networks to be smaller and less centralized than those found in SCI. We did not expect to find a single, large component in the SSCI network, as we found in SCI (Leydesdorff *et al.*, 2013). These expectations have to be conditioned on the fact that 554 journals are fully covered in common by both databases, so some double counting is inevitable.

In addition, one can expect developments such as international collaboration to be very different both in the origins and consequences among the various disciplines in the social sciences (Wagner, 2005). Psychology, for example, can be considered as a "western"-type science more than economics (Bornmann *et al.*, 2012). Furthermore, one can expect that the most-frequently cited papers are unevenly distributed among world regions (Bornmann & Leydesdorff, 2012). In this study, however, we pursue the decomposition in the geographical dimension because the role of national or regional governments under the pressure of internalization is not discipline-specific.

---

[2] The study complements an article about the global map and the descriptive (network) statistics for SCI (Leydesdorff *et al.*, 2013).

After providing maps and descriptive statistics for SSCI (2011) and comparing these with equivalent maps and methodologies for the SCI, we turn to the question of the balance between international and national research agendas: is the research agenda set increasingly by the international network of co-authorship relations and do domestic patterns of co-authorship follow; how may this be different among nations or—in the case of the USA—among states? Using institutional addresses as a next-lower level distribution of the address information, we propose an information-theoretical methodology for addressing this question: is the distribution of internationally co-authored papers over institutions a better predictor of domestically co-authored patterns of co-authorship relations or is the domestic distribution a better predictor of the international one?

Unlike a symmetrical correlation between two distributions, information theory—in this case, Kullback and Leibler's (1951) divergence measure—enables us to distinguish asymmetrically between these two predictions. For reasons of presentation, however, we turn first to the network analysis and the descriptive statistics of SSCI, and only thereafter to the policy-relevant question of how to measure the relative influence of domestic and international networks.

## 2. Methods and data

The entire set of the DVD-version of the *Social Science Citation Index* 2011 (SSCI) was downloaded. This data was brought under the control of relational database management (in the dbf-format using Flagship v7). Initial parsing was complicated by an unexpected finding in the data: we had expected that SSCI online and on the DVD would be identical. However, 32,042 (16.6%) of the 193,338 documents in the set are drawn into the database selectively from SCI using journals that are not otherwise covered by SSCI. (In 2011, 554 journals were fully covered by both databases.) For example, 52 of the 130 papers published in the *Annals of Epidemiology* during 2011 are covered by both indices. In addition to the 2,966 social-science journals covered by the Journal Citation Reports 2011, these additional papers originate from (4,777 − 2,966 =) 1,811 journals that are covered in SCI.[3]

|  | *SSCI* | *SCI* |
|---|---:|---:|
| articles, reviews, and letters | 157,932 | 787,001 |
| a+r+l with addresses | 128,785 | 778,988 |
| addresses | 305,816 | 2,101,384 |
| Authors | 559,321 | 4,660,500 |
| internationally coauthored | 26,667 | 193,216 |
| addresses | 95,442 | 825,664 |
| Authors | 138,063 | 1,401,709 |
| % intern. coauthored | 16.9 | 24.6 |
| % addresses | 31.2 | 39.3 |
| % authors | 24.7 | 30.1 |

**Table 1**: Descriptive statistics of the two databases in 2011.

Table 1 provides the descriptive statistics for the two databases in 2011. Of the 193,338 records in SSCI, 125,385 (64.9%) are articles, 29,748 reviews (15.4%), but only 2,799 (1.4%) are letters.

---
[3] Leydesdorff, Hammarfelt, & Salah (2011) reported that in the Arts & Humanities Citation Index 2008, 3.7% of the 114,929 records were selectively imported from SSCI.

The number of review articles is proportionally much higher in SSCI than in SCI (15.4% versus 3.6%).[4] Only 128,785 (81.5%) of the 157,932 citable items have address information in the bylines, as against 99.0% in SCI. Whereas English is completely dominant in SCI (> 99%), three percent of the papers in SSCI are not in English. German is the second language with 1.3% of the papers, followed by French (0.6%), Spanish (0.3%), and Russian (0.2%) of the papers.

An asymmetrical matrix of documents versus countries was constructed and saved as a systems file in SPSS (v20) for generating relevant statistics. The number of countries was consolidated in SSCI as 187 (versus 201 in SSCI). A number of small countries (e.g., Andorra, Antigua & Barbuda, etc.) were not represented in the SSCI, and one country name ("British West Indies") in the SSCI is obsolete (these islands are now called "West Indies Associated States"). Samoa, Central African Republic, and Equatorial Guinea were not connected, but all the other (183) countries are connected in a single (largest) component of international co-authorship relations.

UCINet (v6.461) was used to generate a symmetrical co-authorship matrix among countries (after changing all values to binary) where a record with three addresses in country A and two addresses in country B is counted as a single relation between these two countries. (An affiliations routine in social-network analysis would otherwise count this as 3 x 2 = 6 relations.) The papers were also fractionally counted: fractional counting means the attribution of each address to a paper in proportion to the number of addresses provided in the byline of the article.

---

[4] "In the *JCR* system any article containing more than 100 references is coded as a review. Articles in 'review' sections of research or clinical journals are also coded as reviews, as are articles whose titles contain the word 'review' or 'overview.'" At http://thomsonreuters.com/products_services/science/free/essays/impact_factor/ (retrieved April 8, 2012).

For example, if two of the three addresses are in country A, the paper is attributed for $2/3^{rd}$ to country A and for $1/3^{rd}$ to country B.

## 3. Results

*3.1 Comparison between SSCI and SCI at the global level*

We generated an interactive map of these relations that was brought online at http://www.leydesdorff.net/intcoll/sosci.htm . This map includes countries with more than 100 citable items in the set (fractionally-counted), and a threshold was set for showing a link at > 100 (integer-counted) coauthored papers. The map is comparable to a map for the SCI 2011 at http://www.leydesdorff.net/intcoll/intcoll.htm , but in this latter case the thresholds were set at 500 (fractionally counted) papers at the nodes and a minimum weight of 500 for the links. For example, the *N* of coauthored publications between Australia and New Zealand is 161 in SSCI (indicated on Figure 2), as against 805 in SCI. Such a ratio of 1:5 is approximately proportional to the difference in the number of records in the two databases.

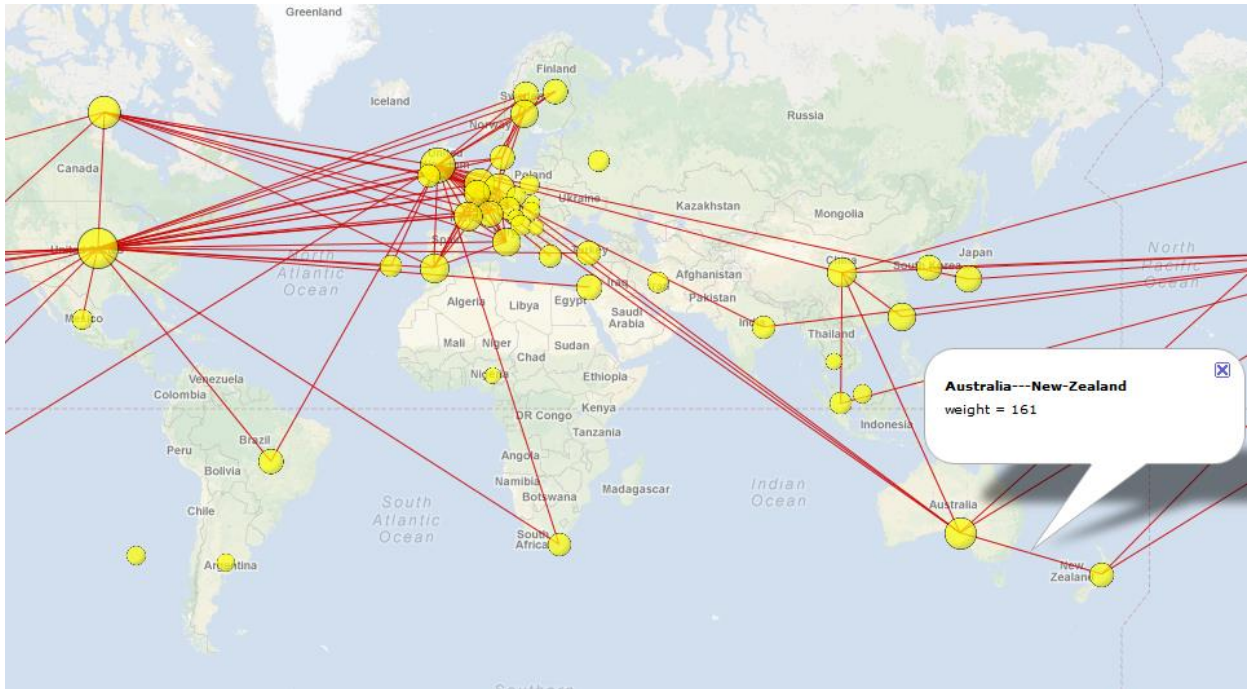

**Figure 2**: International co-authorship network for the SSCI 2011 for countries with more than 100 citable items in the file and for links with a value > 100; at http://www.leydesdorff.net/intcoll/sosci.htm . The size of the nodes is proportionate to the logarithm of the fractionally counted numbers of documents.

In summary, the differences between SSCI and SCI are not so obvious on visual inspection except in terms of the sizes of the two databases. UCINet provides the so-called Quadratic Assignment Procedure or QAP-correlation that enables us to compare the two data-matrices. The QAP between these two matrices is 0.905 ($p < .001$; $N = 201$); but this Pearson-correlation-based measure is inflated because of the large numbers of zeros in the matrices (Ahlgren *et al*., 2003; cf. Egghe & Leydesdorff, 2009). The Jaccard index between the two matrices is only 0.42 which indicates that there are also important differences.

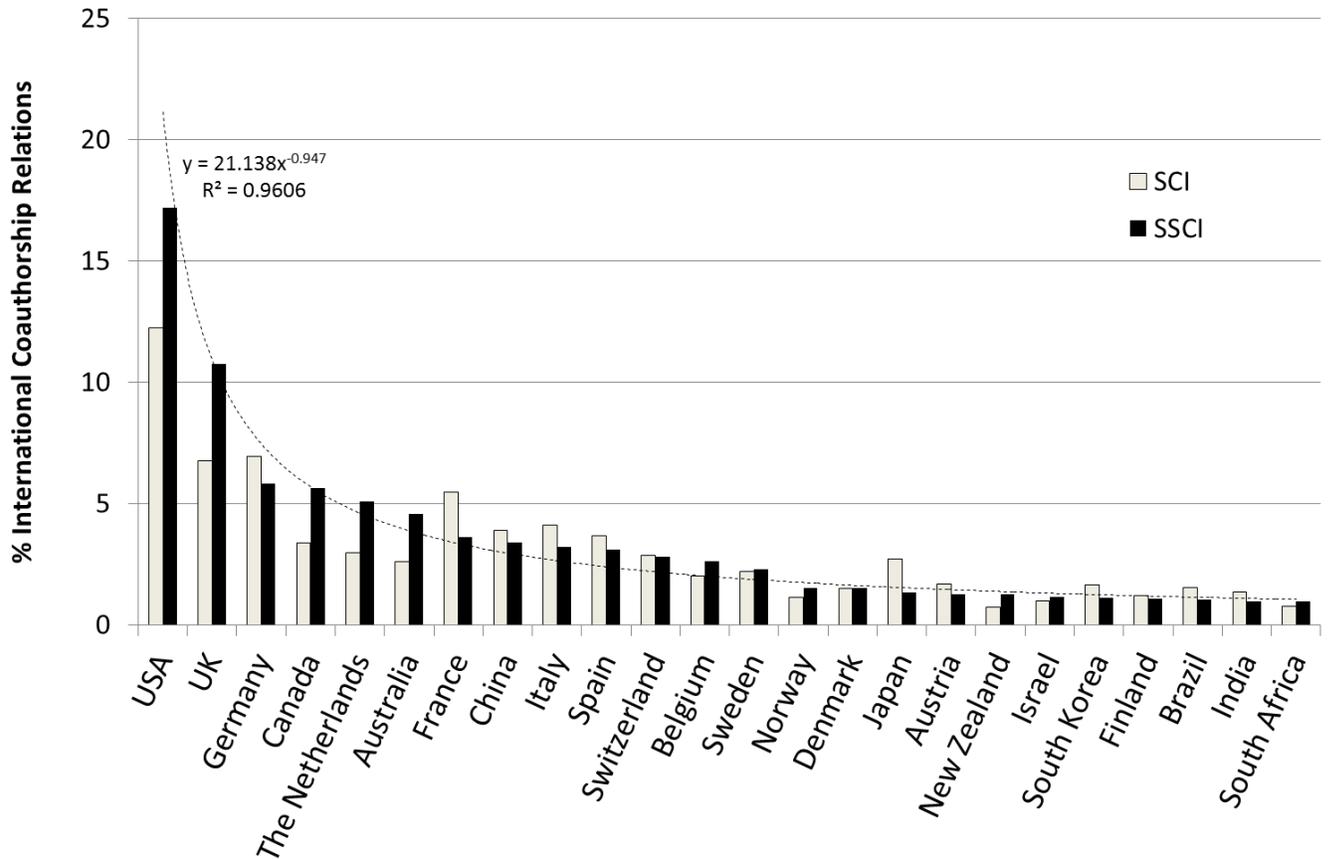

**Figure 3:** Shares of leading countries (≥ 1%) of international co-authorship relations in the Social Science Citation Index 2011, compared with shares in SCI.

Figure 3 shows the percentages of international co-authorship relations in SSCI and SCI for leading nations. The USA, UK, Canada, the Netherlands, and Australia are much more strongly represented in SSCI than SCI. Inversely, France and Japan, and to a lesser extent also Germany, China, Italy, and Spain, are weaker in SSCI than in SCI. The differences are minor for the Scandinavian countries. We tested these differences as proportions against each other at the level of countries using the *z*-test for two independent proportions; Table 2 provides the results for the top-groups in terms of relative prominence in SSCI and SCI, respectively.

| z-test | SSCI | | SCI |
|---|---|---|---|
| USA | 54.5 | Russia | -38.6 |
| UK | 52.2 | Japan | -27.1 |
| Canada | 38.9 | France | -26.8 |
| Netherlands | 38.2 | Czech Republic | -25.5 |
| Australia | 37.6 | Armenia | -21.1 |
| Lebanon | 23.3 | Serbia | -21.0 |
| Nigeria | 21.7 | Byelarus | -20.8 |
| Dominican Rep. | 19.3 | Poland | -20.5 |
| Kenya | 18.4 | Georgia | -19.4 |
| New Zealand | 18.2 | Ukraine | -18.3 |
| Uganda | 17.4 | Slovakia | -16.9 |
| Singapore | 15.7 | Egypt | -15.7 |
| Belgium | 13.2 | Germany | -14.8 |
| . . . | | . . . | |

**Table 2:** Rank ordering of *z*-scores for nations comparing percentage contributions to international co-authorship relations in SSCI and SCI 2011, respectively.

In SSCI, the above noted first five countries lead the list, but some African countries and countries with relatively small numbers are also part of the list. (Perhaps this is an effect of former students and postdocs returning to their home countries.) A negative score in Table 2 means that the proportion in the SSCI is lower than in the SCI. (All numbers in Table 2 are highly significant [$p<.001$].[5]) On the reverse list indicating a relative prominence in SCI, countries that previously belonged to the sphere of influence of the Soviet Union dominate, but Japan and France are also high. The prominence of the English-speaking countries raises questions about the language bias in the choices of journals. As noted, only 3% of the documents included in SSCI were in languages other than English notwithstanding the drastic expansion of the regional coverage of SSCI during the last years (Testa, 2010).

---

[5] These z-scores are all highly significant, but the significance levels can be Bonferroni-adjusted because of family-wise correction of possible Type-I errors. In the case of approximately 200 comparisons $\alpha = p/200 = 0.05/200 = 0.00025$. The z-values (> 4) in Table 3 are significant at this level.

|                                    | *SSCI* | *SCI* |
|------------------------------------|--------|-------|
| N of nodes                         | 187    | 201   |
| largest component                  | 183    | 200   |
| N of lines                         | 2,695  | 5,981 |
| Density                            | 0.154  | 0.293 |
| Average degree                     | 28.82  | 58.89 |
| Network Betweenness Centralization | 0.217  | 0.101 |
| Communities (Blondel *et al.*, 2008) | 8    | 5     |
| Network Clustering coefficient     | 0.524  | 0.614 |

**Table 3**: Network statistics and parameters compared for the SCI and SSCI 2011.

Table 3 shows the comparison of a number of network statistics for the two networks. The density of the network in the social sciences is approximately half of that in SCI, and the average degree is also half. This can partly be an effect of the size differences between the two databases, but centrality is not a valued measure. Betweenness centralization, however, is twice as high at the network level, indicating that the leading nations are more central in this network than in SCI. Although the clustering coefficients do not vary widely, the number of communities (using the Louvain algorithm in Pajek; Blondel *et al.*, 2008) seems considerably larger in SSCI than in SCI. However, removal of the isolates reduces the number of communities to four in both databases.

In summary, the results of the comparison suggest that the two networks are not so different at the level of the observable values except of a size effect between the two databases. However, centralization seems more pronounced in SSCI than SCI. Centralization in the SSCI is most likely due to the fact that advanced countries have more to spend on social sciences, or to an English language bias, or to both.

*3.2    Centralization in SSCI*

Table 4 shows that the (global) network in SSCI is completely dominated by the same group of countries in all respects, although the rank-order may differ a bit among these nations. In addition to the five countries already mentioned, France, Germany, Switzerland, and Italy are also central to the network. At lower levels of the hierarchy the ordering is less stable, with Spearman rank correlations of the order of .8 among the distributions of the centrality measures.

| Country     | Degree | Country     | Betweenness | Country     | Closeness |
|-------------|--------|-------------|-------------|-------------|-----------|
| USA         | 156    | USA         | 0.221       | USA         | 0.856     |
| UK          | 137    | UK          | 0.100       | UK          | 0.785     |
| France      | 113    | France      | 0.088       | France      | 0.710     |
| Netherlands | 110    | Netherlands | 0.047       | Netherlands | 0.701     |
| Canada      | 109    | Australia   | 0.044       | Canada      | 0.698     |
| Germany     | 108    | Germany     | 0.042       | Germany     | 0.696     |
| Australia   | 104    | Canada      | 0.038       | Australia   | 0.685     |
| Switzerland | 103    | Switzerland | 0.030       | Switzerland | 0.682     |
| Italy       | 94     | Italy       | 0.024       | Italy       | 0.660     |
| Belgium     | 93     | Spain       | 0.021       | Belgium     | 0.657     |

**Table 4**: Top-10 countries in terms of degree, betweenness, and closeness centrality in the global network based on SSCI 2011.

As noted, four communities were indicated in the giant component (183 of the 187 countries identified) using the Louvain algorithm (Blondel *et al.*, 2008) in Pajek. Figure 4 shows these four communities in the network using the hundred most-frequent lines as a threshold.

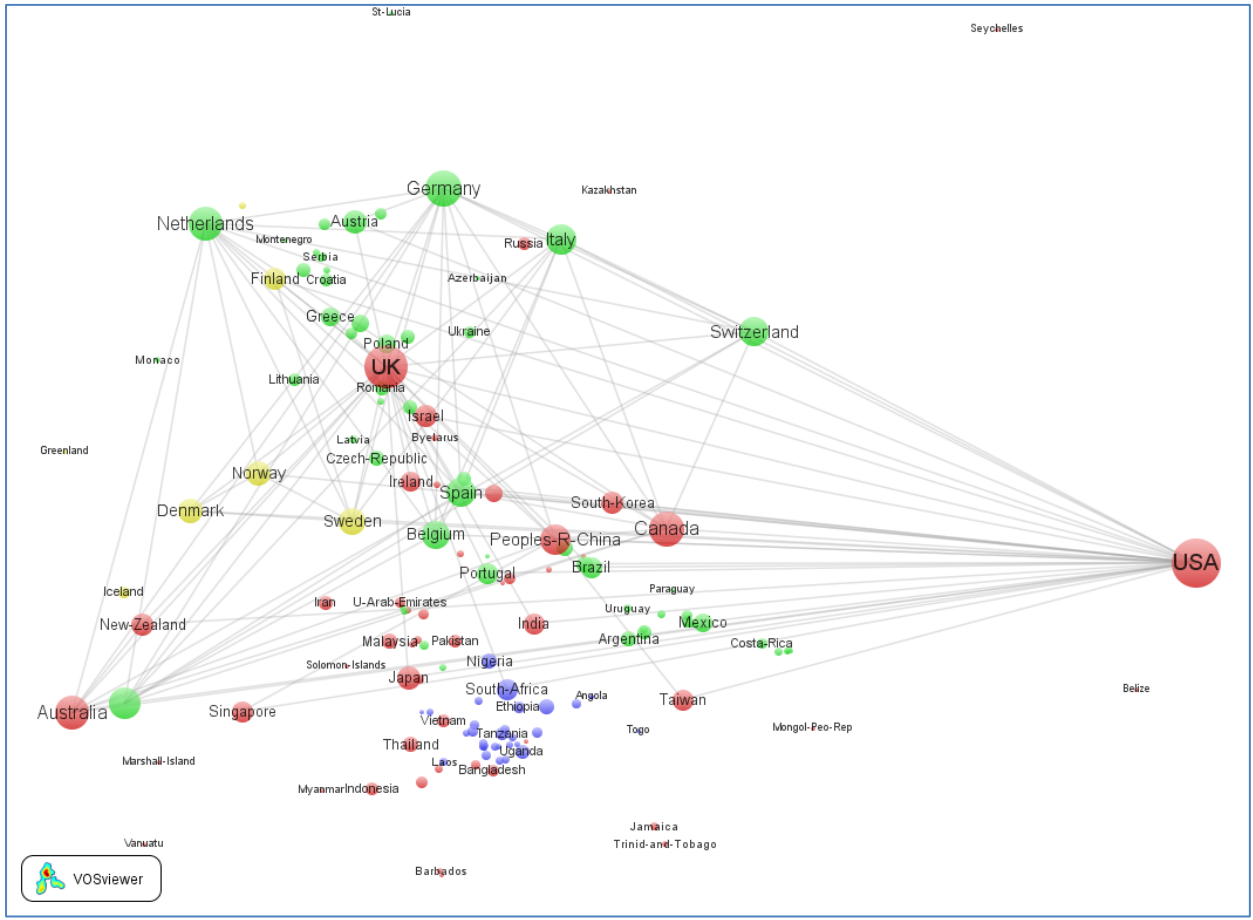

**Figure 4**: Four communities of 183 nations using international co-authorship relations in SSCI 2011; hundred most-frequent lines shown; VOSViewer was used for the mapping (Van Eck & Waltman, 2010); $Q = 0.12$ (Blondel *et al*., 2008).

The four communities distinguished in SSCI (Figure 4) are: (*i*) an Anglo-Saxon group to which the Asian-Pacific nations are also attached (red); (*ii*) a European (continental) group to which most Eastern-European nations (except for Russia) are attached, with also the Latin-American countries (green); (*iii*) a Scandinavian group (yellow); and (*iv*) an African community of nations (blue). In some runs, France and some former French colonies appear as a separate (francophone) community, but most of the time France is part of the European group albeit in an outlier position (next to "Australia" and therefore without a label in Figure 4; cf. Van Eck & Waltman, 2010).

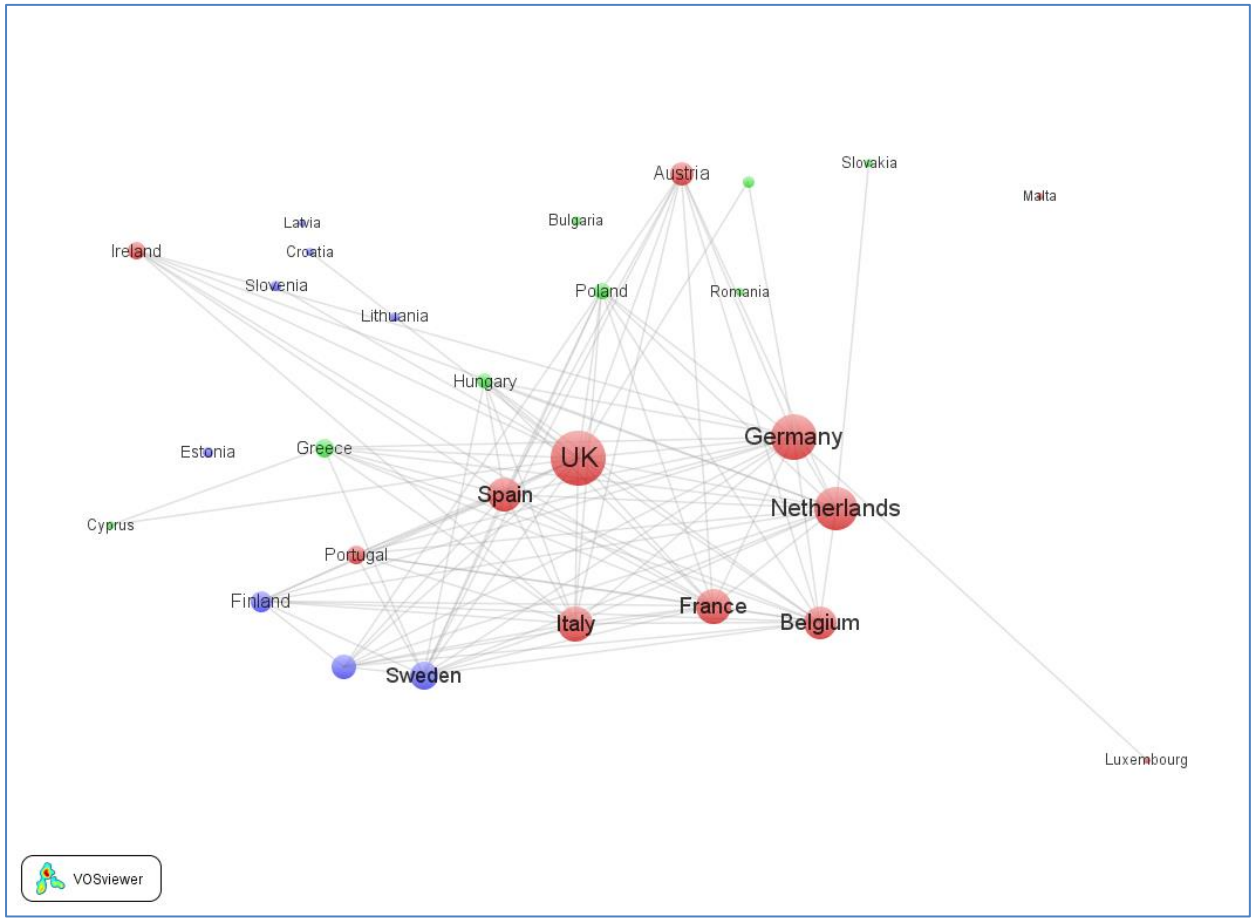

**Figure 5**: The EU-28 (including Croatia); 100 most-frequent links penciled in; VOSViewer used for the visualization; three communities with modularity $Q = 0.36$ (Blondel *et al.*, 2008).

Figure 5 shows the decomposition of the 28 EU member states that were divided in a continental versus (UK + Ireland) grouping in Figure 4. The grouping is now different and more pronounced since the modularity is much higher ($Q = 0.36$). The first and largest group contains eleven of the fifteen nations of the old EU-15. The Scandinavian countries (Denmark, Sweden, and Finland) form a separate group. Greece is classified (with Cyprus) among the member states that have joined the EU since 2004. Most of these latter states connect to the central countries, but much less so among themselves.

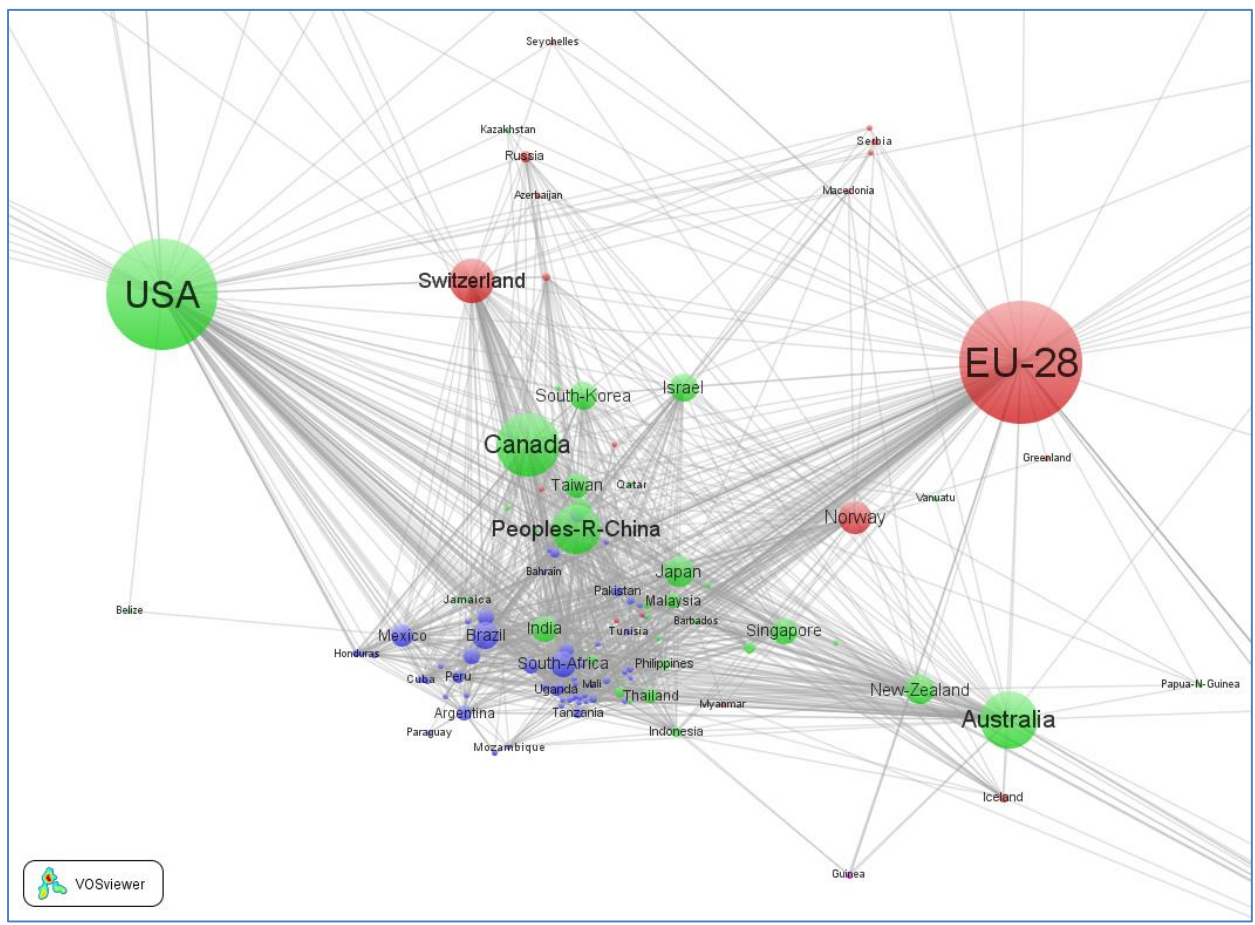

**Figure 6**: The global network of 156 countries (= 187 − 28 + 1) based on combination of the member-states of the EU-28. (All links shown; VOSViewer used for the visualization; $Q = 0.12$.)

In Figure 6, the 28 EU member states are collapsed into a single node at the global level (using Pajek). This figure shows a bi-polar structure between the USA and the EU-28 at the global level. The Asian-Pacific group is spanned between the USA and Australia as leading nations and includes China, India, and Japan. European non-member states side with the EU-28. Certain other countries are positioned far to the left of (and exclusively related to) the USA (e.g., the Bahamas), or to the right of the EU-28 depending, for example, on previous colonial relations (e.g., Netherlands Antilles). This result suggests that competition between the USA and the EU has become more important than collaboration at the world level. As noted, this is very different when the EU is decomposed into its member states.

*3.3 Ego-networks of individual nations*

The matrix of co-authorship relations among 187 countries allows the user to draw sub-maps such as the one of the set of nearest neighbours of a single nation. For this purpose, the matrix is available online in the Pajek format at http://www.leydesdorff.net/intcoll/sosci.net .

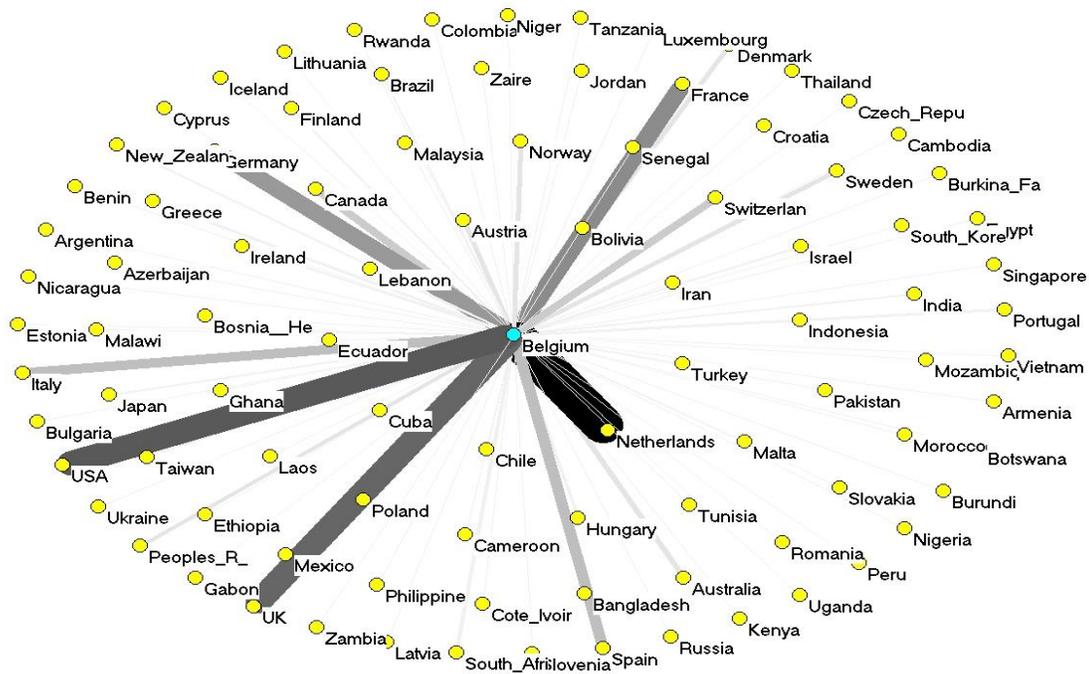

**Figure** 7: 3,583 international co-authorship relations with authors in 93 other countries on the basis of 2,229 documents with an Belgian address in the SSCI 2011; *k*=1 network in Pajek.

Figure 7, for example, shows the *Ego*-network of Belgium in this network. The Netherlands is the main partner of Belgium, followed by the USA, the UK, and the other neighboring countries: Germany and France. Relations with industrially advanced countries dominate the picture.

Similar maps can be made for any nation in the set following the procedure (in Pajek) outlined in an Appendix.[6]

## 4. Policy implications

In our opinion, the policy question behind the study of international collaborations is whether and how these developments at the global level affect the effectiveness of science policies by national or regional governments. If these networks develop internationally, does this lead to unintended spill-overs of national investments and thus confine the influence of (national) governments setting priorities in funding specific disciplinary areas or interdisciplinary topics (Park & Leydesdorff, 2013). The relations between international and domestic networks can change the dynamics in the competition among nation states. Is the network already so internationalized that one could, for example, also act as a brinkman without increasing R&D funding further because others would fill in the gaps?

These issues are currently important for the legitimacy of R&D funding in a period in which on the one side one expects growth from science and technology, and on the other is under pressure to control budgets. The European Commission (2000), for example, has pleaded since the "Lisbon Treaty" for increasing national R&D budgets to 3% of GDP, but more recently this objective was abandoned in favor of austerity policies aimed at reducing government spending. Although the social sciences are not viewed as strategically linked to national industrial development and competitiveness to the same degree as the natural sciences and engineering, the same question can be asked around the social sciences as the natural sciences. Yet, policy considerations could be different. For example, one would expect that the social sciences are less internationalized than the natural sciences.

---

[6] A similar file for the Science Citation Index is available at http://www.leydesdorff.net/intcoll/sci.net .

What can the distributions teach us in this respect? Is international co-authoring more important than domestic co-authorship, or are national systems still firmly in place (Skolnikoff, 1993)? How can one make the above results relevant for this debate? Our data contain both domestic and international (co-)authorship data in terms of addresses. Table 5, for example, ranks universities and other research organizations contained in the file in terms of their contribution to the aggregate of citable items in SSCI.

| (a) | Citable items (b) | Percentage domestic addresses (c) | Percentage internationally coauthored (d) | Country (e) |
|---|---|---|---|---|
| HARVARD-UNIV | 3104 | 72.1 | 27.9 | USA |
| UNIV-TORONTO | 2260 | 61.7 | 38.3 | CANADA |
| UNIV-MICHIGAN | 2166 | 84.6 | 15.4 | USA |
| UNIV-N-CAROLINA | 2011 | 82.7 | 17.3 | USA |
| COLUMBIA-UNIV | 1698 | 72.0 | 28.0 | USA |
| UNIV-CALIF-LOS-ANGELES | 1687 | 77.9 | 22.1 | USA |
| UNIV-WASHINGTON | 1677 | 80.9 | 19.1 | USA |
| UNIV-PENN | 1668 | 80.7 | 19.3 | USA |
| UNIV-ILLINOIS | 1607 | 82.9 | 17.1 | USA |
| UNIV-PITTSBURGH | 1500 | 83.1 | 16.9 | USA |
| YALE-UNIV | 1466 | 77.4 | 22.6 | USA |
| UNIV-MINNESOTA | 1429 | 81.7 | 18.3 | USA |
| DUKE-UNIV | 1360 | 74.9 | 25.1 | USA |
| UNIV-BRITISH-COLUMBIA | 1300 | 59.8 | 40.2 | CANADA |
| UNIV-WISCONSIN | 1297 | 81.3 | 18.7 | USA |
| UNIV-CALIF-SAN-FRANCISCO | 1261 | 79.4 | 20.6 | USA |
| UCL | 1244 | 55.7 | 44.3 | UK |
| KINGS-COLL-LONDON | 1183 | 59.8 | 40.2 | UK |
| UNIV-SYDNEY | 1179 | 67.2 | 32.8 | AUSTRALIA |
| MCGILL-UNIV | 1146 | 57.9 | 42.1 | CANADA |
| STANFORD-UNIV | 1144 | 71.5 | 28.5 | USA |
| UNIV-MARYLAND | 1131 | 80.6 | 19.4 | USA |
| UNIV-MELBOURNE | 1101 | 61.1 | 38.9 | AUSTRALIA |
| NYU | 1100 | 77.0 | 23.0 | USA |
| PENN-STATE-UNIV | 1090 | 81.5 | 18.5 | USA |
| NORTHWESTERN-UNIV | 1086 | 79.1 | 20.9 | USA |
| UNIV-AMSTERDAM | 1080 | 60.9 | 39.1 | NETHERLANDS |
| UNIV-QUEENSLAND | 1061 | 67.0 | 33.0 | AUSTRALIA |
| UNIV-OXFORD | 1044 | 49.2 | 50.8 | UK |
| JOHNS-HOPKINS-UNIV | 999 | 76.3 | 23.7 | USA |
| .... | .... | ... | ... | ... |
| Totals (of 36,725 addresses) | 304,840 | 210,376 | 94,464 | |
| 2,866 organizations with more than 10 citable items | 247,172 80.8% | 171,653 81.6% | 75,489 79.9% | |

**Table 5**: Thirty universities with 1,000 or more citable items in the SSCI 2011.

Of the 305,816 addresses attributed at the level of the file (see Table 1), 304,840 (99.7%) can be included in the analysis using the first subfield of the address information in SSCI. However, the resulting 36,725 organizations include different name variants, typos, and sometimes sub-organizations. In order to remove this noise from the data, we used only organizations with more than 10 citable items in the file.[7]

We focused on the 2,866 organizations with more than 10 citable items in the file. This is 7.8% of all organizations in the file, but these organizations published more than 80% of the citable documents, both domestically and internationally. The file also contains city-addresses, and it would be possible to disaggregate further; for example, the "Chinese Academy of Science" (500 citable items) into its different affiliations. However, in this study we are not specifically interested in the domestic networks per se (Leydesdorff & Persson, 2010), but rather in the relation between domestic and international addresses.[8]

---

[7] This number was 2,420 in the analysis of the SCI using 100 instead of 10 citable items as the threshold (Wagner *et al.*, in preparation).

The columns (c) and (d) in Table 5 correlate significantly ($p<.01$; $N = 2,702$),[9] but are also different: Pearson's $r = 0.842$ and Spearman's $\rho = 0.612$. Our interest, however, is in the difference: to what extent is the layer of internationally coauthored papers driving the domestic publication pattern at the institute level?

4.1 *Using Kullback-Leibler divergence as an asymmetrical test*

As against correlation measures, Kullback & Leibler's (1951) divergence provides an asymmetrical measure of the difference between two distributions that allows us to specify the extent to which the one distribution can be considered as a predictor of the other or *vice versa*. This measure can be derived from Shannon's entropy measures (Theil, 1972; Leydesdorff, 1990, 1991), and informs us (non-parametrically) about the difference between two distributions in bits of information. In formula format, one can write the measure as follows:

$$I_{q|p} = \sum_i q_i * log_2(q_i/p_i)$$

In this formula $I_{q|p}$ measures the information in the message that the *a priori* (independent) probability distribution ($\Sigma_i\ p_i$) was changed into the *a posteriori* (dependent) one ($\Sigma_i\ q_i$). If the two distributions are similar, knowledge about the one distribution will predict the other perfectly and no information is generated using this measure (because the log[1] = 0). However, if one probability distribution is different from the other, the Kullback-Leibler divergence is always larger than zero (Theil, 1972, pp. 59 ff.). In other words, the better predictor variable is the one that generates the least information upon the measurement of the *a posteriori* distribution.

---

[8] One exception was made for a record with "Harvard-Univ, Sch Med, Vancouver, BC, Canada" as address that is relabeled with the US-address in Cambridge MA because otherwise the analysis would be disturbed for incidental reasons.
[9] $N = 2,702$ of the 2,866 organizations (94.4%) have non-zero values in both columns.

| (a) | Citable items of 66 countries (b) | All records (c) |
|---|---|---|
| N of organizations | 2,866 | 36,725 |
| with (non-zero) both domestic and international publications | 2,702 (94.3%) | 6,914 (18.8%) |
| $I(dom|int)$ in mbits | 347.5 | 368.5 |
| $I(int|dom)$ in mbits | 364.6 | 392.7 |

**Table 6**: Data and results for the Kullback-Leibler divergence test.

Let us apply this measure to the question of the extent to which the domestic distribution of publications in SSCI 2011 can be used as a predictor of the distribution of internationally coauthored publications at the level of nations, and *vice versa*. Institutions in 87 nations (of the 187) passed the threshold of having more than 10 citable items in the file and were included in this analysis. However, 21 of these nations are represented with only a single institutional address with more than 10 citable items. In the case of a single unit of analysis, the distribution does not contain uncertainty. Therefore, we use only (87 − 21 =) 66 countries for the comparison in Table 6. The international distribution is a slightly better predictor of the national one, both within the restricted set of 66 countries (column b in Table 6) and across all records in the file (column c).

Using precisely the same procedure, one can analogously decompose the USA into 52 states (including Puerto Rico), but only (52 – 2 =) states[10] are attributed with more than 10 citable items in the database and with more than a single address above this threshold (Table 7). Also in this case, the international distribution of addresses proves, at the aggregated level to be a marginally better predictor of the domestic one than is the case *vice versa*. This is a notable result because both Figure 1 and Table 5 above showed that domestic co-authorship is still larger than international co-authorship in the USA.

| (a) | 50 states of the USA (b) | All records with US addresses (c) |
|---|---|---|
| N of addresses | 904 | 10,610 |
| with non-zero both domestic and international publications | 806 (89.2%) | 1,716 (16.2%) |
| $I$(dom|int) in mbits | 150.3 | 175.6 |
| $I$(int|dom) in mbits | 154.4 | 195.8 |

**Table 7**: Data and results for 50 US states; analogously to data and results in Table 6.

*4.2  Decomposition in terms of nations and US-states*

Using the distributions at the institutional level, both countries and states can be evaluated at the aggregated level in terms of the question of whether $I$(international|domestic) is larger or smaller than $I$(domestic|international). In the maps of nations and states we color those units white in

---

[10] Puerto Rico (PR) and Wyoming (WY) were excluded given these thresholds. In Puerto Rico, only the University of Puerto Rico had more than ten (i.e., 74) publications in SSCI 2011, of which 62 were internationally coauthored. The University of Wyoming had 118 publications in 2011, of which 18 were internationally co-authored.

which the international dimension provides the better predictor, whereas the others are colored blue.

Note that the values of Kullback-Leibler divergence are quantitative (in bits of information), but we use the measure here as a binary test because normalization issues are involved in comparisons across countries or states. In the case of a country with only two addresses, for example, changing a single publication may lead to a different result. However, elaboration of these normalization issues would lead us beyond the scope of the current study, so we decided to use the measure only as a binary test of whether the domestic or the international distribution provides a better prediction at the national level.[11]

4.2.1. *Countries*

Of the 66 countries, the domestic pattern is the better prediction in 30 cases, and in the remaining 36 cases the international pattern provides the better predictor. (These numbers were 26 and 34, respectively, using the data of SCI; Wagner *et al.*, in preparation.) The results are visualized in Figure 8.

---

[11] Were the two predictions equal, we would count this conservatively as a "domestically" driven unit of analysis. However, this case did not occur in our data.

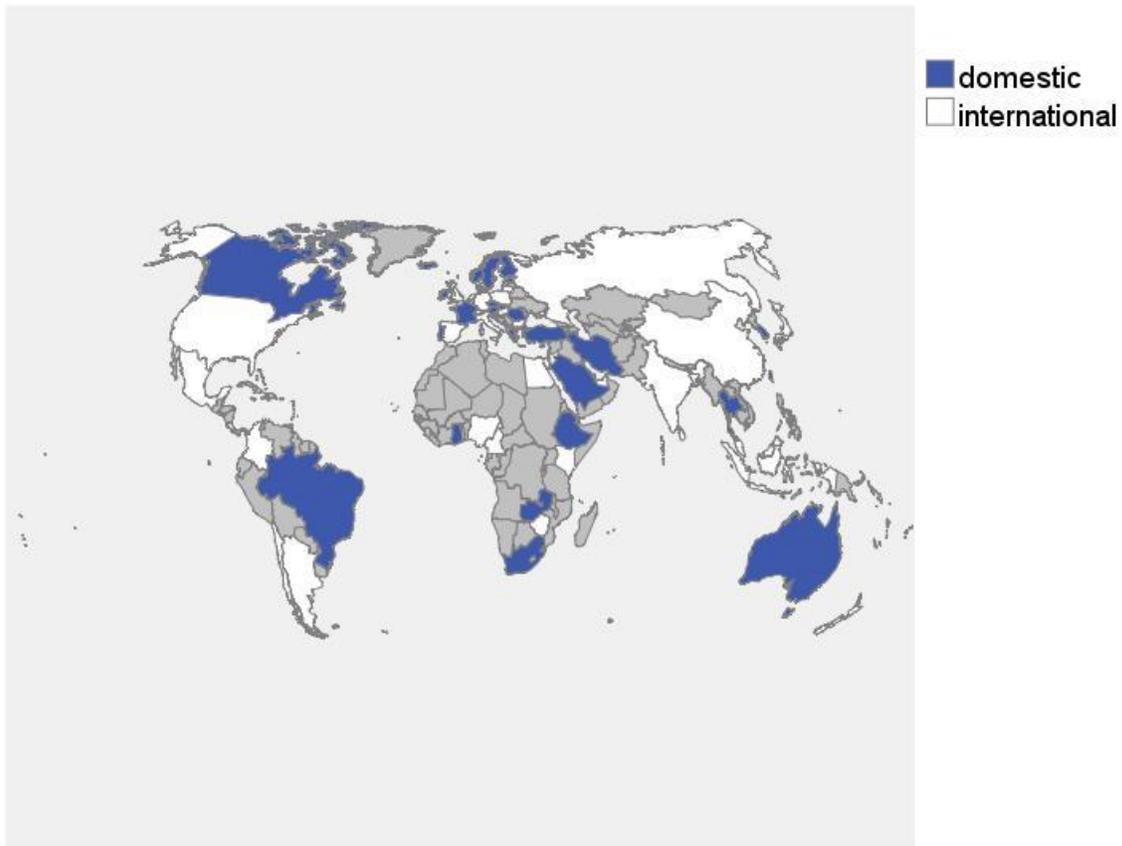

**Figure 8**: Map of 66 countries in terms of the domestic distribution (blue) or the international distribution (white) as the better predictor.

Although the UK itself is shown as oriented internationally more than domestically, some of the large countries of the Commonwealth (Australia, Canada, South-Africa) are more domestically oriented on this indicator.

*4.2.2   Decomposition of the USA in terms of states*

Figure 9 follows up with a map of the states of the USA. The distribution of domestically (co)authored papers is the better predictor of the internationally co-authored ones in 33 cases (blue in Figure 2), whereas in 17 states the situation is the reverse (white). (These numbers were 18 and 28 in SCI, respectively.) Using USA data, therefore, domestic patterns are on average the better predictor in SSCI, but not in SCI. This confirms the expectation.

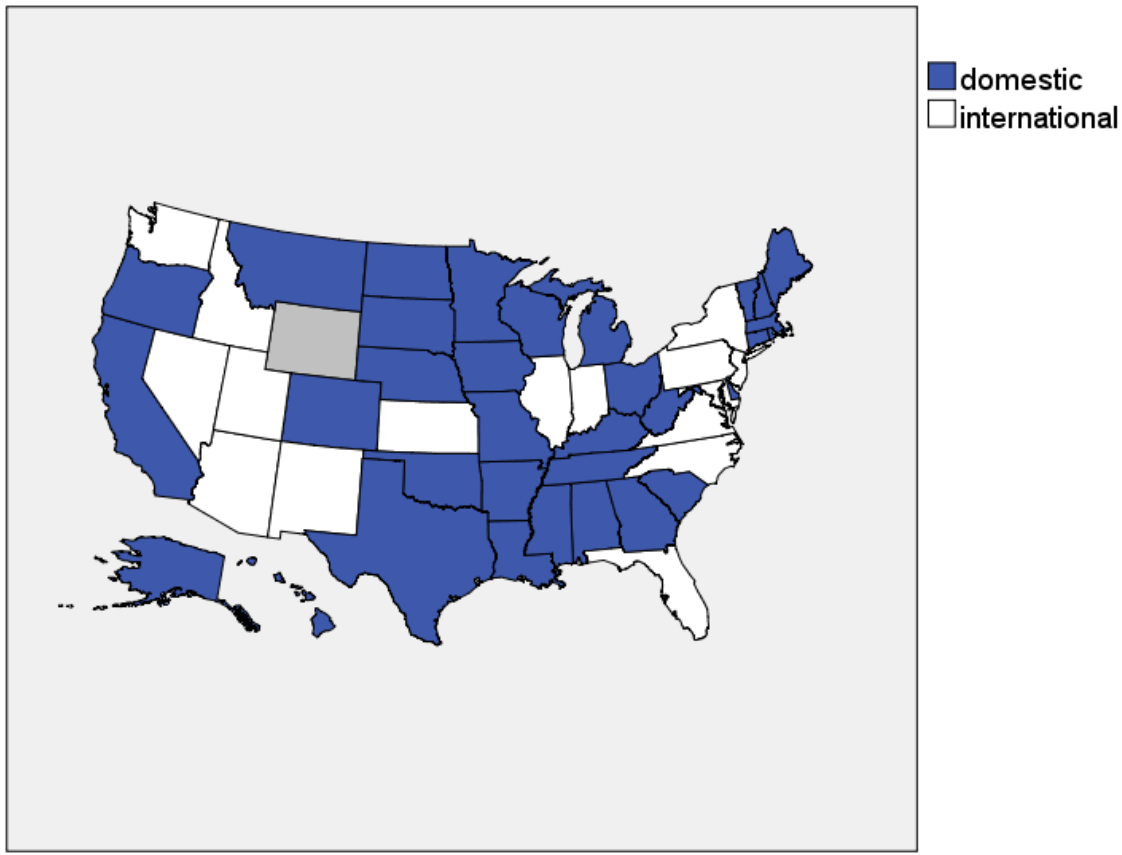

**Figure 9**: Map of the states of the USA in terms of the domestic distribution (blue) or the international distribution (white) as the better predictor. (The value of Wyoming is missing.)

In summary, in terms of numbers of nations (both in SSCI and SCI), the distributions of internationally co-authored papers (over institutions) predicts the distribution of domestically co-authored papers better than *vice versa*. At the level of states of the USA, however, the domestic pattern prevails at the national level in SSCI. Of the 47 states that we were able to compare between SSCI and SCI, 26 had the same orientation in terms of this test.

*4.2.3 European Union*

Figure 10 shows the same measure applied to the countries of the EU-28. Among these member-states, Latvia is the only country that has no institutions with ten or more publications included in SSCI 2011, and therefore Latvia was excluded from this analysis. In France and the Scandinavian countries, the domestic co-authorship relations are the better predictors, but in the other core countries such as the UK, Germany, Italy, Spain, and the Netherlands the international patterns prevail in prediction. As noted, the European program incentivize international collaboration, but this policy incentive does not have to lead to international co-authorship in the output.

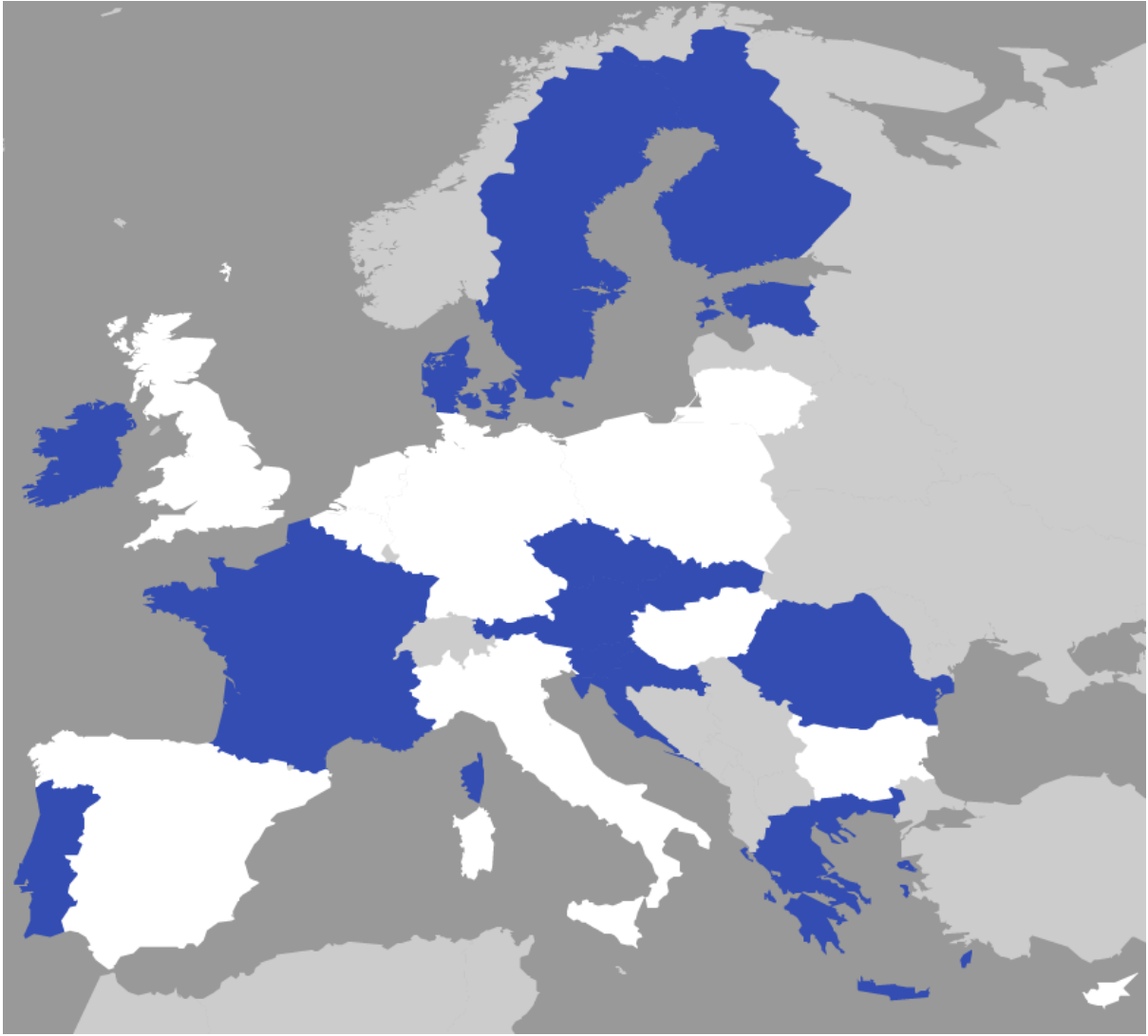

**Figure 10**: European Union in terms of the domestic distribution (blue) or the international distribution (white) as the better predictor.

We emphasize that using the Kullback-Leibler divergence measure for this test is crude and can only be considered as a first attempt to make the statistical analysis perhaps more relevant for this policy debate. The descriptive statistics cannot inform the political discourse about the relative importance of the international network for determining agendas. However, we have used only a single year, and the results may be sensitive to parameter choices such as thresholds. But we wished to convey the message that the policy dilemma is not an either-or choice, but a trade-off: some nations and states have become internationalized more than others. In accordance with Figure 1 (that opened this paper), one should not underestimate domestic co-authorship as important for the system, but an important layer of internationally collaboration was added in recent decades.

## 5. Conclusion

Using international co-authorship relations in SSCI, we found four groups of nations at the global level that are most connected: (*i*) an Asian-Pacific and Anglosaxon one, including also the UK and Ireland; (*ii*) a continental European one, also related to Latin America; (*iii*) the Scandinavian countries as a separate group; and (*iv*) a group of African nations, including South-Africa. Within the EU-28 (including Croatia), a core group of eleven nations are distinguished from the Scandinavian nations (Denmark, Sweden, and Finland) on the one hand, and the accession countries on the other (Greece is part of the latter group along with Cyprus.) Collapsing the EU-28 into a single unit leads to a bi-polar view of the world with all the European nations on one side, and the USA and the Asian-Pacific nations on the other. The less-developed countries of Africa and Latin America are now grouped separately.

The descriptive comparison at the aggregate level did not show convincing differences between SSCI and SCI except for the size differences between the two databases, which are reflected, for example, in the densities of the networks. One should, however, keep in mind that the coverage of the social sciences in SSCI is different from that of the natural and life sciences in SCI because publication outlets in national languages may not or only very partially be covered by SSCI (Larivière *et al.*, 2006; Ossenblok *et al.*, 2012 and in press). One expects cultural and linguistic influences to be larger in the social sciences than in the natural sciences and engineering (Nederhof, 2006). In other words, we have mainly the internationalized layer of the social sciences in view. Yet, the results teach us that internationalization is as important in this internationally oriented layer of the social sciences just as in the natural and life sciences, and that similar dynamics are operating. The specific position of the Scandinavian nations in SSCI, furthermore, seems robust (at different levels of aggregation) and therefore perhaps worth further investigation.

As a second objective of the study, we raised the question of what these descriptive statistics teach us with reference to the policy-analytical issue of whether and how internationalization makes it necessary to rethink assumptions of national policies. The network perspective makes accessibility to knowledge at the international level perhaps more important than the size of the participation. Although these two aspects may be not independent, other competencies are required. In a network environment, for example, one may be able to excel with a relatively small investment; large investments may lead to unintended spill-overs.

We were able to quantify that there is a balance between organization at the national (or state) level in terms of institutions and self-organization among scholars at the international level. The Kullback-Leibler divergence measure enables us to use the international and domestic distributions of co-authorship relations as relative predictors of each other. As was perhaps to be expected, we found a prevalence of the international dimension in some countries and (US) states and in others a prevalence of the domestic dimension. Given the static nature of our (one-year) data, however, there is little that we can say other than that internationalization seems still to have momentum. An extension of this study to more than a single year (2011) would provide further research perspectives.

Apart from these empirical results, the operationalization teaches us that there is no once-and-for-all answer to this relation. Countries and states have progressed differently in terms of internationalization. However, the test at the aggregated level suggests that the international dimension prevails as a predictor of the domestic one and not the reverse. This indicates that the cybernetic expectation that the next-order level is constructed bottom-up, but tends to take over top-down control, may already have passed a point of no return.

**Appendix**

Generation of a country's ego-network from the file SoSCI.net at

http://www.leydesdorff.net/intcoll/sosci.net, using Pajek v3.

The subsequent steps for drawin a map like Figure 7 are as follows:

- Download the data from http://www.leydesdorff.net/intcoll/sosci.net

- Read this network into Pajek (File > Network > Read: "sosci.net");

- Network > Create Partition > *k*-neighbours; select node number and distance 1.

- Operations > Network + Partition > extract subnetwork 0-1; "0" for ego, "1" for *k*=1 neighbours;

- Partition > Make Cluster > Vertices from selected cluster > 1 (only *k*=1 neighbours)

- Operations > Network + Partition > Transform > Remove Lines > Inside Cluster 1 (that is, links among the k=1 neighbours)

- Draw > Network + first partition

One may have to turn on Labeling under Options after drawing the figure.

**Acknowledgement**
We thank Thomson Reuters for access to the data; some of us acknowledge support from the SSK (Social Science Korea) Program funded by National Research Foundation of South Korea; NRF-2010-330-B00232. We thank two anonymous referees for their comments.